\documentclass[10pt,twocolumn]{article}
\usepackage{latexsym,graphicx}
\usepackage{latexsym,graphicx}
\newcommand{\be}{\begin{equation}}
\newcommand{\ee}{\end{equation}}
\def\n{\noindent}
\catcode `\@=11 \catcode `\@=12
\begin{document}
\begin{center}
\large{\bf {DARK ENERGY MODELS WITH VARIABLE EQUATION OF STATE PARAMETER}} \\
\vspace{10mm}
\normalsize{Anil Kumar Yadav\footnote{corresponding author}, Farook  Rahaman$^2$ and Saibal Ray$^3$}\\
\vspace{4mm}
\normalsize{$^1$Department of Physics, Anand Engineering
College, Keetham, Agra -282 007, India} \\
\normalsize{E-mail: abanilyadav@yahoo.co.in, akyadav@imsc.res.in}\\
\vspace{2mm}
\normalsize{$^2$Department of Mathematics, Jadavpur University, Kolkata - 700 032, India}\\
\normalsize{E-mail: farook\_rahaman@yahoo.com}\\
\vspace{2mm}
\normalsize{$^3$Department of physics, Government College of Engineering and Ceramic Technology, Kolkata 700 010,
India}\\
\normalsize{E-mail: saibal@iucaa.ernet.in}
\end{center}
\vspace{10mm}
\begin{abstract}
\n The dark energy models with variable equation of state parameter $\omega$ are investigated
by using law of variation of Hubble's parameter that yields the constant value of deceleration
parameter. Here the equation of state parameter $\omega$ is found to be time dependent and its existing range
for this model is consistent with the recent observations of SN Ia data, SN Ia data (with CMBR anisotropy)
and galaxy clustering statistics. The physical significance of the dark energy models have also been discussed.
\end{abstract}
\\
\\
\\
\\

\smallskip
\n Keywords :  LRS Bianchi type V Universe, Dark Energy, EoS Parameter\\
\section{Introduction}
Observations on large scale structure (LSS) \cite{ref1}  and cosmic microwave background
radiation \cite{ref2}$-$\cite{ref6}
indicate that the Universe is highly homogeneous and isotropic on large scales. Under standard
assumptions on the matter content, in general relativistic cosmological models, isotropy is a
special feature, requiring a high degree of fine tuning in order to reproduce the observed Universe.
Recent years have witnessed the emergence of the idea of an accelerating Universe due to some observational results
\cite{ref7}$-$\cite{ref10}.
This signifies a remarkable shift in cosmological research from expanding Universe to accelerated expanding
Universe. Now, the problem lies in detecting an exotic type of unknown repulsive force, termed as dark energy~(DE)
which is responsible for the accelerating phase of the Universe.
The detection of DE would be a new clue to an old puzzle: the gravitational effect of the zero-point
energies of particles and fields \cite{ref11}. The total with other energies, that are close to homogeneous
and nearly independent of time, acts as DE. The paramount characteristic of the DE is a
constant or slightly changing energy density as the Universe expands, but we do not know the nature of
DE very well (see \cite{ref12}$-$\cite{ref20} for reviews on DE). DE has been conventionally
characterized by the equation of state (EoS) parameter $\omega=p/\rho$ which is not necessarily constant.
The simplest DE candidate is the vacuum energy $(\omega = -1)$, which is argued to be equivalent to the
cosmological constant $(\Lambda)$ \cite{ref21}. The other conventional alternatives, which can be described by minimally
coupled scalar fields, are quintessence $(\omega>-1)$, phantom energy $(\omega<-1)$ and quintom
(that can across from phantom region to quintessence region) as evolved and have time dependent EoS parameter.
Some other limits obtained from observational results coming from SN Ia data \cite{ref22} and SN Ia data
collaborated with CMBR anisotropy and galaxy clustering statistics \cite{ref23} are $-1.67<\omega<-0.62$ and
$-1.33<\omega<-0.79$ respectively. However, it is not at all obligatory to use a constant value of $\omega$.
Due to lack of observational evidence in making a distinction between constant and variable $\omega$, usually
the equation of state parameter is considered as a constant \cite{ref24,ref25} with phase wise value $-1$, $0$,
$1/3$ and $+1$ for vacuum fluid, dust fluid, radiation and stiff fluid dominated universe, respectively.
But in general, $\omega$ is a function of time or redshift \cite{ref26}$-$\cite{ref28}. For instance, quintessence
models involving scalar fields give rise to time dependent EoS parameter $\omega$ \cite{ref29}$-$\cite{ref32}.
Also some literature is available on models with varying fields, such as cosmological model with variable equation
of state parameter in Kaluza-Klein metric and wormholes \cite{ref33,ref34}.
In recent years various form of time dependent $\omega$ have been used for variable $\Lambda$ models \cite{ref35,ref36}.
Recently Ray et al \cite{ref37}, Akarsu and Kilinc \cite{ref38} have studied variable EoS parameter for
generalized dark energy model.

Bianchi type-V universe is generalization of the open universe in FRW cosmology and hence its study is important in
the study of DE models in a universe with non-zero curvature \cite{ref39}. A number of authors such as
Collins \cite{ref40}, Maartens and Nel \cite{ref41}, Wrainwright et al \cite{ref42}, Canci et al \cite{ref43},
Pradhan et al \cite{ref44} and Yadav \cite{ref45} have studied Bianchi type-V model in different physical contexts.
Recently Yadav and Yadav \cite{ref46} have studied anisotropic DE models with variable EoS parameter. In this paper,
we have investigated the DE
models with variable $\omega$ in Bianchi type-V Universe. This paper is organized as follows:
The metric and field equation are presented in section 2.
In section 3, we deal with the solution of field equations and discussion. Finally the result are
discussed in section 4.\\

\section{The Metric and Field  Equations}
We consider LRS Bianchi type V metric in the form
\begin{equation}
\label{eq1}
ds^2 = -dt^2 +A^{2}dx^2 +B^{2} e^{2x} \left(dy^2 + dz^2\right)
\end{equation}
where A and B are the function of t only.\\
The simplest generalisation of EoS parameter of perfect fluid may be to determine
the EoS parameter separately on each spatial axis by preserving the diagonal form
of the energy-momentum tensor in a consistent way with the considered metric.
Thus, the energy momentum tensor of fluid is taken as
\begin{equation}
\label{eq2}
T_{i}^{j} = diag\left[T_{0}^{0}, T_{1}^{1}, T_{2}^{2}, T_{3}^{3}\right]
\end{equation}
Then, one may parametrize it as follows,
\[
 T_{i}^{j} = diag\left[\rho, -p_{x}, -p_{y}, -p_{z}\right]
= diag\left[1, -\omega_{x}, -\omega_{y}, -\omega_{z}\right]\rho
\]
\begin{equation}
\label{eq3}
= diag\left[1, -\omega, -\left(\omega + \delta \right), -\left(\omega + \delta\right)\right]\rho
\end{equation}
where $ \rho $ is the energy density of fluid, $p_{x}$, $p_{y}$ and $p_{z}$ are the pressures and $\omega_{x}$,
$\omega_{y}$ and $\omega_{z}$ are the directional EoS parameters along the x, y and z axes respectively.
$\omega$ is the derivation-free EoS parameter of the fluid. We have parametrized the deviation from isotropy
by setting $\omega_{x} = \omega$ and then introducing skewness parameter $\delta$ that are deviation from
$\omega$ along y and z axis respectively.\\
The Einstein field equations, in gravitational units ($8\pi G =1$ and $c=1$), are
\begin{equation}
\label{eq4}
R_{ij} - \frac{1}{2}Rg_{ij} = -T_{ij}
\end{equation}
where the symbols have their usual meaning.\\
The Einstein's field equation (\ref{eq4}) for the Bianchi type V space-time (\ref{eq1}),
in case of (\ref{eq3}), lead to the following system of equations
\begin{equation}
\label{eq5}
2\frac{B_{44}}{B} +\frac{B_{4}^2}{B^2} - \frac{3}{A^2} = \rho
\end{equation}
\begin{equation}
\label{eq6}
2\frac{B_{44}}{B} + \frac{B_{4}^2}{B^2} - \frac{1}{A^2} = -\omega\rho
\end{equation}
\begin{equation}
\label{eq7}
\frac{A_{44}}{A} + \frac{B_{44}}{B} + \frac{A_{4}B_{4}}{AB} -\frac{1}{A^2} = -\left(\omega + \delta\right)\rho
\end{equation}
\begin{equation}
\label{eq8}
\frac{A_{4}}{A} - \frac{B_{4}}{B} = 0
\end{equation}
Here, the sub indices 4 in A, B, and elsewhere denote differentiation with respect to t.\\
Integrating equation (\ref{eq8}), we obtain
\begin{equation}
\label{eq9}
A = kB
\end{equation}
where $k$ is the positive constant of integration. We substitute the value of equation (\ref{eq9})
in equation (\ref{eq7}) and subtract the result from equation (\ref{eq6}), we obtain that the skewness
parameter on z-axis is null i.e.
$$\delta = 0$$
Thus system of equations from (\ref{eq5}) - (\ref{eq8}) may be reduce to
\begin{equation}
\label{eq10}
2\frac{B_{44}}{B} +\frac{B_{4}^2}{B^2} - \frac{3}{A^2} = \rho
\end{equation}
\begin{equation}
\label{eq11}
2\frac{B_{44}}{B} + \frac{B_{4}^2}{B^2} - \frac{1}{A^2} = -\omega\rho
\end{equation}
Now we have three linearly independent equations (\ref{eq9}) - (\ref{eq11}) and four unknown
parameters $\left(A, B, \omega, \rho\right) $. Thus one extra condition is needed to solve the system
completely. To do that, we have used the law of variation of Hubble's parameter that yields a constant
value of deceleration parameter.\\
The average scale factor of Bianchi type V metric is given by
\begin{equation}
\label{eq12}
R = \left(AB^{2}e^{2x}\right)^\frac{1}{3}
\end{equation}
We define, the generalised mean Hubble's parameter $H$ as
\begin{equation}
\label{13}
H = \frac{1}{3}\left(H_{1} + H_{2} + H_{3}\right)
\end{equation}
where $H_{1} = \frac{A_{4}}{A}$, $H_{2} = H_{3} = \frac{B_{4}}{B}$ are the directional Hubble's
parameter in the direction of x, y and z respectively.\\
Equation (13), may be reduces to
\begin{equation}
\label{eq14}
H = \frac{R_{4}}{R} = \frac{1}{3}\left(\frac{A_{4}}{A} +2\frac{B_{4}}{B}\right)
\end{equation}
Here the line-element (\ref{eq1}) is completely characterized by Hubble's parameter H.
Therefore, let us consider that mean Hubble parameter H is related to average scale factor $R$
by following relation
\begin{equation}
\label{eq16}
H = k_{1}R^{-n}
\end{equation}
where $k_{1} > 0$ and $n \geq 0$, are constant. Such type of relation have already been considered
by Berman \cite{ref47} for solving FRW models. Later on many authors
(Singh et al \cite{ref48,ref49} and references therein)
have studied flat FRW and Bianchi type models by using the special law of Hubble parameter that yields
constant value of deceleration parameter.\\
The deceleration parameter is defined as
\begin{equation}
\label{eq16}
q = -\frac{RR_{44}}{R_{4}^2}
\end{equation}
From equations (\ref{eq14}) and (15), we get
\begin{equation}
\label{eq17}
R_{4} = k_{1}R^{-n+1}
\end{equation}
\begin{equation}
\label{eq18}
R_{44} = -k_{1}^{2}(n-1)R^{-2n+1}
\end{equation}
Using equations (\ref{eq14}) and (\ref{eq18}), equation (\ref{eq16}) leads to
\begin{equation}
\label{eq19}
q = n-1~~(constant)
\end{equation}
The sign of $q$ indicates whether the model inflates or not. The positive sign of $q$ corresponds to
standard decelerating model where as the negative sign of $q$ indicates inflation. It is remarkable to mention here that
though the current observations of SN Ia (Permutter et al 1999, Riess et al 1998) and CMBR favour accelerating models
i.e. $q < 0$. But both do not altogether rule out the decelerating ones which are also consistent with these
observations (Vishwakarma 2003).\\
From equation (\ref{eq17}), we obtain the law of average scale factor R as
\begin{equation}
\label{eq20}
R =  \left[ \begin{array}{ll}
            \left(Dt + c_{1}\right)^\frac{1}{n}  & \mbox { when $n \neq 0$}\\
 c_{2}e^{k_{1}t}                                      & \mbox { when $n = 0$}
            \end{array} \right.
\end{equation}
where $c_{1}$ and $c_{2}$ are the constant of integration.\\
From equation (\ref{eq20}), for $n \neq 0$, it is clear that the condition for accelerating  expansion
of Universe is $0< n < 1$.\\
\section{Solutions of the Field  Equations and Discussion}
\subsection{Case~(i): when $n\neq0$}
Equations (\ref{eq8}), (\ref{eq14}) and (\ref{eq20}) lead to
\begin{equation}
\label{eq21}
B = l\left(Dt + c_{1}\right)^{\frac{1}{n}}
\end{equation}
From equations (\ref{eq9}) and (\ref{eq21}), we obtain

\begin{equation}
\label{eq22}
A = L\left(Dt + c_{1}\right)^{\frac{1}{n}}
\end{equation}
where l is the constant of integration and $ L = kl $.\\
Thus the Hubble's parameter $(H)$, scalar of expansion $(\theta)$, shear scalar $(\sigma^2)$ and spatial
volume $(V)$ are given by
\begin{equation}
\label{eq23}
H = \frac{k_{1}}{(Dt+c_{1})}
\end{equation}
\begin{equation}
\label{eq24}
\theta = 3H = \frac{3k_{1}}{(Dt+c_{1})}
\end{equation}
\begin{equation}
\label{eq25}
\sigma^2 = \frac{3k_{1}^2 -D^2}{(Dt+c_{1})^2}
\end{equation}
\begin{equation}
\label{eq26}
V = Ll^{2}(Dt+c_{1})^{\frac{3}{n}}e^{2x}
\end{equation}
Using equations (\ref{eq10}), (\ref{eq20}) and (\ref{eq21}), the energy density of the
fluid is obtained as
\begin{equation}
\label{eq27}
\rho = \frac{3D^2}{n^2\left(Dt+c_{1}\right)^2} - \frac{3}{L^2\left(Dt+c_{1}\right)^\frac{2}{n}}
\end{equation}
Using equations (\ref{eq11}), (\ref{eq21}), (\ref{eq22}) and (\ref{eq27}), the equation of state
parameter $\omega$ is obtained as
\begin{equation}
\label{eq28}
\omega = \frac{\left[\frac{n^2}{L^2D^2}\left(Dt+c_{1}\right)^\frac{2(n-1)}{n}+2n-3\right]}
{3\left[1-\frac{n^2}{L^2D^2}\left(Dt+c_{1}\right)^\frac{2(n-1)}{n}\right]}
\end{equation}
\begin{figure}
\begin{center}
\includegraphics[width=3.0in]{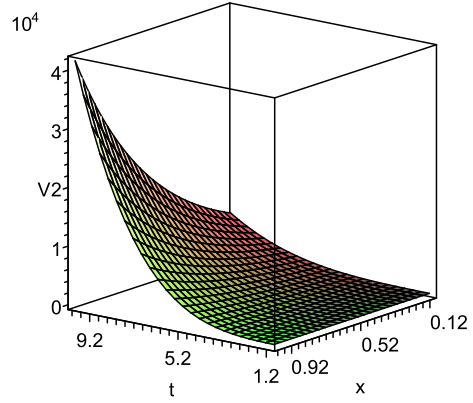}
\caption{The plot of spatial volume V for n = .8}
\label{fg:anil267fig1.eps}
\end{center}
\end{figure}
From equation (\ref{eq27}), we note that $\rho(t)$ is the decreasing function of time.
This behaviour is clearly shown in Fig. 2, as a representative case with appropriate choice
of constants of integration and other physical parameters using reasonably well known situations.
From equation (\ref{eq28}), it is observed that the equation of state parameter $\omega$ is time dependent,
it can be function of redshift z or scale factor R as well. The redshift dependence of $\omega$ can be linear
like $\omega(z)=\omega_{0}+\omega^{1}z$ with $\omega^{1} = \left(\frac{d\omega}{dz}\right)_{z}=0$ \cite{ref50,ref51} or
nonlinear as $\omega(z)=\omega_{0}+\frac{\omega_{1}z}{1+z}$ \cite{ref52,ref53}. The SN Ia data suggests
that $-1.67<\omega<-0.62$ \cite{ref22} while the limit imposed on $\omega$ by a combination
of SN Ia data (with CMB anisotropy) and galaxy clustering statistics is $-1.33<\omega<-0.79$ \cite{ref23}.
So, if the present work is compared with experimental results mentioned above then, one can conclude that
the limit of $\omega$ provided by equation (\ref{eq28}) may accommodated with the acceptable range of EoS parameter.
The value $\omega =-1$ is the case of vacuum fluid dominated Universe. But here, we are dealing with the solution for
$n\neq0$, therefore in this case, vacuum fluid dominated universe is meaningless.
Also we see that during evolution of Universe, at an instant of time $t=\frac{1}{D}
\left(\left[\frac{(3-2n)L^2D^2}{n^2}\right]^\frac{n}{2n-2}-c_{1}\right)$, the $\omega$ vanishes. Thus at
this particular time, our model represents dusty Universe.\\

\n The critical density $(\rho_{c})$ and density parameter $(\Omega)$ are given by
\begin{equation}
\label{eq29}
\rho_{c} = \frac{3k_{1}^2}{(Dt+c_{1})^2}
\end{equation}
\begin{equation}
\label{eq30} \Omega=\frac{D^2L^2 -n^2
(Dt+c_{1})^{\frac{2n-2}{n}}}{k_{1}^2L^2n^2}
\end{equation}

\begin{figure}
\begin{center}
\includegraphics[width=3.0in]{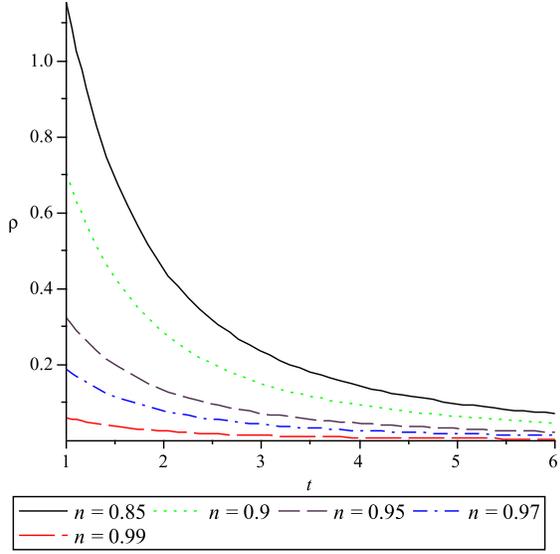}
\caption{The plot of energy density ($\rho$) vs. time (t)}
\label{fg:anil267fig2.eps}
\end{center}
\end{figure}

\begin{figure}
\begin{center}
\includegraphics[width=3.0in]{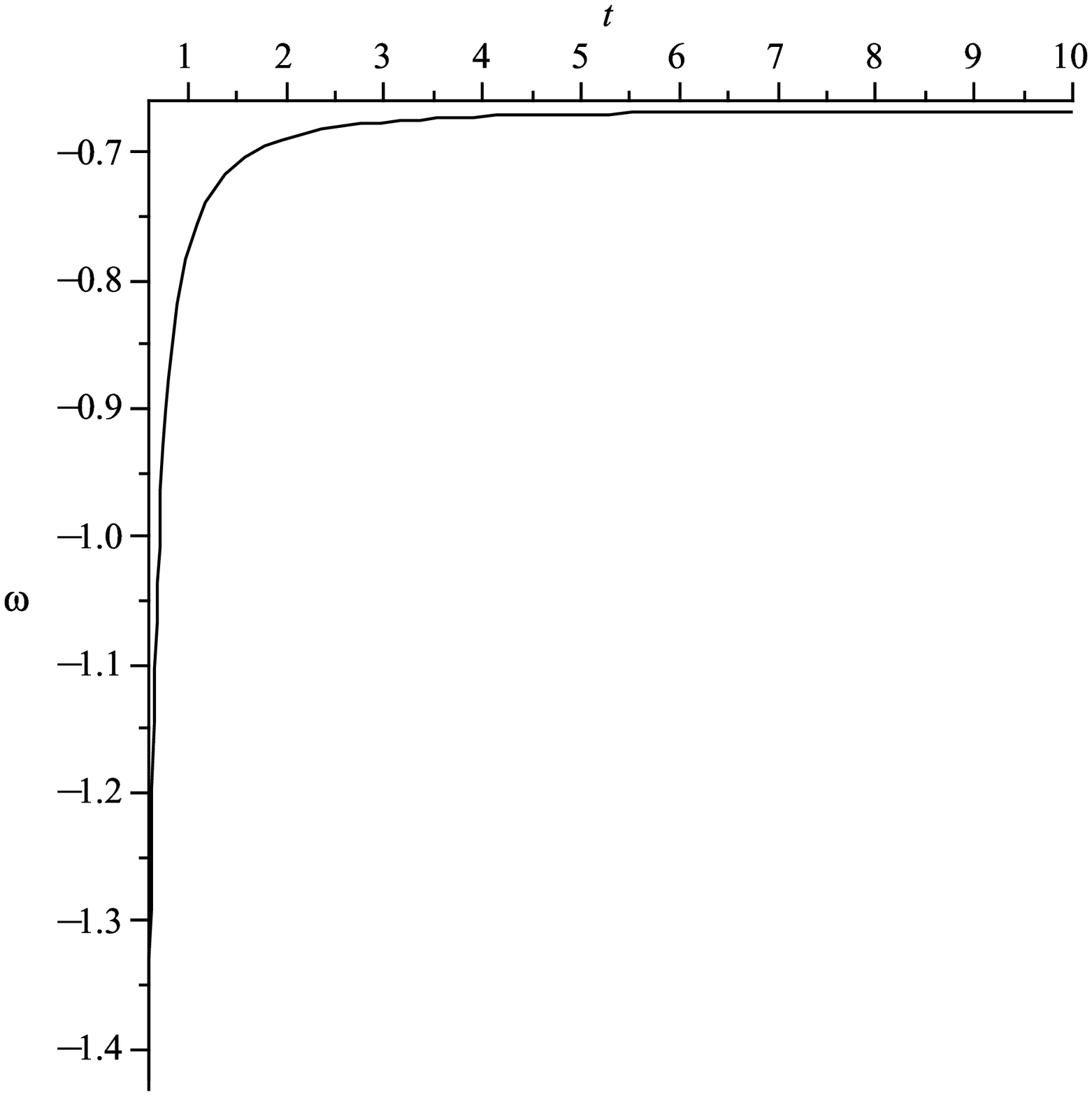}
\caption{The plot of EoS parameter ($\omega$) vs. time (t) for $n
= 0.5$} \label{fg:anil267fig3.eps}
\end{center}
\end{figure}

\begin{figure}
\begin{center}
\includegraphics[width=3.0in]{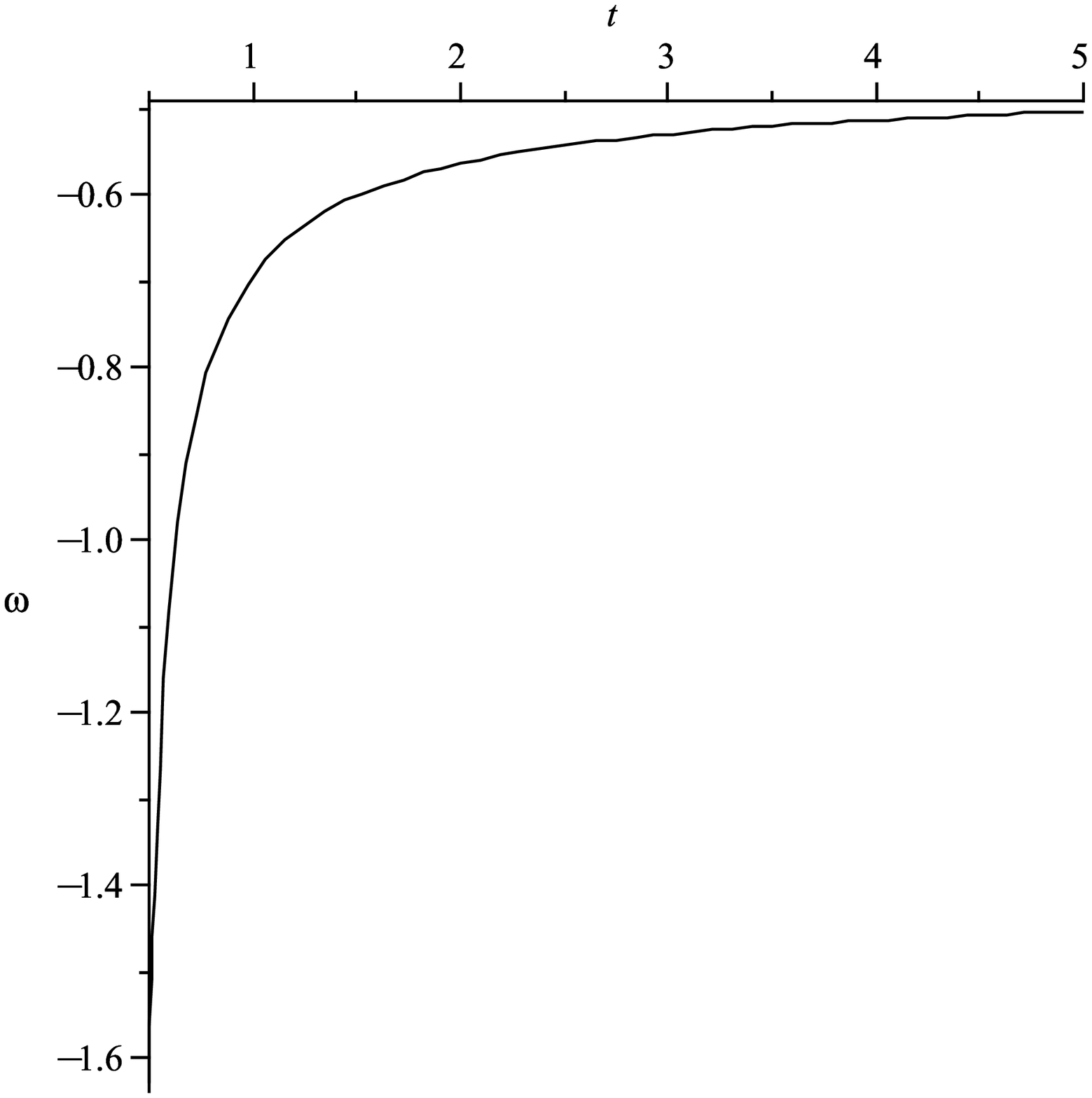}
\caption{The plot of EoS parameter ($\omega$) vs. time (t) for
$n  = 0.85$} \label{fg:anil267fig4.eps}
\end{center}
\end{figure}

\begin{figure}
\begin{center}
\includegraphics[width=3.0in]{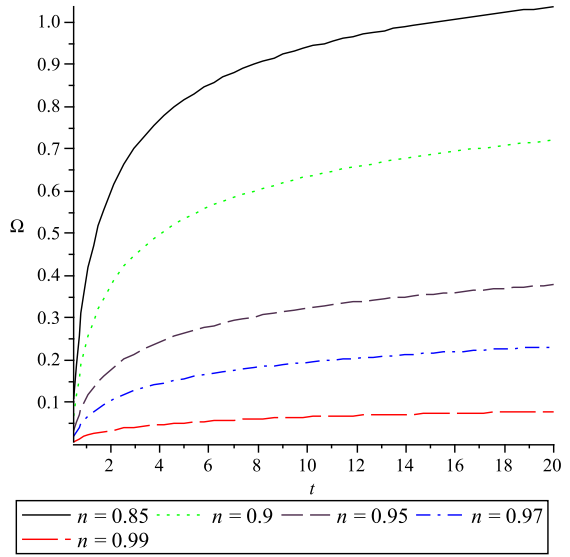}
\caption{The plot of density parameter ($\Omega$) vs. time (t)
for various value of n}. \label{fg:anil27fig5.eps}
\end{center}
\end{figure}

\n The variation of equation of state parameter $(\omega)$ with
cosmic time (t) is  depicted in figures~ 3 and 4 as a
representative case with appropriate choice of constants. Fig. 3
and 4, clearly show that $\omega$ is evolving with negative value
and the existing range of $\omega$ is in nice agreement with SN
Ia data \cite{ref22}. Thus our model is a realistic model. The
plots of $\omega$ for n=0.5 and n=0.85 indicate that $\omega$
merge well with SN Ia and CMBR observations [Fig. 3 and Fig. 4].
The variation of density parameter $(\Omega)$ with time in
accelerating mode
of Universe is clearly shown in Fig. 5.\\
\begin{figure}
\begin{center}
\includegraphics[width=3.0in]{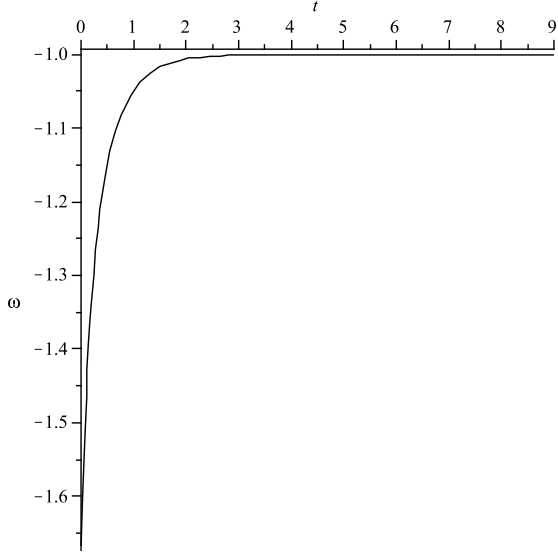}
\caption{The plot of EoS parameter ($\omega$) vs. time (t) for
$n=0$}. \label{fg:anil27fig6.eps}
\end{center}
\end{figure}
\subsection{Case(ii): when $n=0$}
Equations (\ref{eq8}), (15) and (\ref{eq21}) lead to
\begin{equation}
\label{eq31}
B=l_{0}e^{k_{1}t}
\end{equation}
From equations (\ref{eq9}) and (\ref{eq31}), we obtain
\begin{equation}
\label{eq32}
A=L_{0}e^{k_{1}t}
\end{equation}
where $\l_{0}$ is constant of integration and $L_{0}=kl_{0}$.\\
Thus the Hubble's parameter $(H)$, scalar of expansion $(\theta)$, shear scalar $(\sigma^2)$, and spatial volume $(V)$
are given by
\begin{equation}
\label{eq33}
H = k_{1}
\end{equation}
\begin{equation}
\label{eq34}
\theta = 3k_{1}
\end{equation}
\begin{equation}
\label{eq35}
\sigma^2 = 2k_{1}^2
\end{equation}
\begin{equation}
\label{eq36}
V = L_{0}l_{0}^{2}e^{(3k_{1}t+2x)}
\end{equation}
Using equations (\ref{eq10}), (\ref{eq31}) and (\ref{eq32}), the energy density of the
fluid is obtained as
\begin{equation}
\label{eq37}
\rho=\frac{3\left(k_{1}^2L_{0}^2-e^{-2k_{1}t}\right)}{L_{0}^2}
\end{equation}
Using equations (\ref{eq11}), (\ref{eq31}), (\ref{eq32}) and (\ref{eq37}), the equation of state
parameter $\omega$ is obtained as
\begin{equation}
\label{eq38}
\omega=\frac{\left(e^{-2k_{1}t}-3k_{1}^2L_{0}^2\right)}{3\left(k_{1}^2L_{0}^2-e^{-2k_{1}t}\right)}
\end{equation}
The variation of the EoS parameter $\omega$ with cosmic time (t) is shown in Fig. 6. The value of
$\omega$ is found to be negative which is supported by SN Ia data and galaxy clustering statistics \cite{ref22,ref23}.

\n The critical density $(\rho_{c})$ and density parameter $(\Omega)$ are given by
\begin{equation}
\label{eq39}
\rho_{c} = 3k_{1}^2
\end{equation}
\begin{equation}
\label{eq40}
\Omega = \frac{k_{1}^2L_{0}^2 - e^{-2k_{1}t}}{L_{0}^2k_{1}^2}
\end{equation}

\section{Concluding Remarks}
In this paper, we have studied dark energy model with variable EoS parameter $\omega$. The model
is derived by using law of variation of Hubble's parameter that yields a constant value of deceleration
parameter. We have studied two cases, $(3.1)$ and $(3.2)$ for $n\neq0$ and $n=0$ respectively. In both the
cases, $\omega$ is found to be time varying and negative which is consistent with recent observations \cite{ref22,ref23}.
In case (i), we have shown that, $\omega$ is evolving from $\omega <-1$ and finally end up with $\omega > -1$,
representing two phases of universe i.e. $\omega < -1$ (Phantom fluid dominated universe) and  $\omega >-1$
(quintessence),[Fig. 3 and Fig. 4]. Where as in case (ii), for $n=0$, from
equation (\ref{eq38}), we have obtained that,
at cosmic time $t=0$, $\omega<-1$ and when $t\rightarrow\infty$, $\omega\rightarrow-1$.
Therefore, one can conclude that at early stage, universe was dominated by phantom fluid and at late time
it will be vacuum fluid dominated universe [Fig. 6]. Also it is seen that that in both cases, $\frac{\sigma}{\theta} =
constant $, therefore the proposed models do not approach to isotropy at any time. \\

\n Here the age of universe is given by
$$T_{0} = \left(\frac{1}{k_{1}}\right)H_{0}^{-1}$$
which is also differ from the present estimate i.e. $T_{0} = H_{0}^{-1} \approx 14Gyr$. But if we take,
$k_{1} = 1$ then our model $(n\neq 0)$ is in good agreement with present age of universe.\\

\n The main fetures of the models are as follows\\

\n \textbullet~~ Though there are many candidates such as cosmological constant, vacuum energy, space-time curvature,
cosmological nuclear energy, etc as reported in the vast literature for DE, the proposed models in this paper
at least present a new candidate (EoS parameter) as possible suspect of the dark energy.\\

\n \textbullet~~ The dark energy models are based on exact solution of Einstien's field equations for the Bianchi
type V space-time filled with perfect fluid. To our knowledge, the literature hardly witnessed this sort of exact solution
for Bianchi type V space-time. So the derived DE models add one more feather to the literature. The DE
models presents the dynamics of EoS parameter $\omega$ provided by equations (\ref{eq28}) and (\ref{eq38})
may accommodated with the
acceptable range $-1.67<\omega<-0.62$ of SN Ia data (Knop et al 2003).

\section*{Acknowledgements}
One of Authors (AKY) would like to thank The Institute of Mathematical Science (IMSc), Chennai, India
for providing facility and support where part of this work was carried out.


\end{document}